# Birefringent Rydberg Dark Matter from Cosmic Dust

*Keith Johnson, Massachusetts Institute of Technology, Cambridge, MA 02139*

Water nanoclusters ejected to space from abundant amorphous water-ice-coated cosmic dust have been proposed to constitute Rydberg baryonic dark matter [1]. This phenomenological model is now qualitatively supported by several recent observations: (1) the laboratory-generated ejection of water nanoclusters from amorphous water-ice; (2) the absence of dark matter in Galaxy NGC 1052-DF2; (3) a recent study of dark matter in distant galaxies reporting the direct interaction of dark matter with galactic baryonic matter; (4) the bullet cluster; and (5) the birefringence of the cosmic microwave background. A link between Rydberg dark matter and dark energy and its relevance to "quintessence" are reviewed. The use of the James Webb Space Telescope to test this model further is suggested.

## I. INTRODUCTION

In a recent publication [1], it has been conjectured that stable water nanoclusters are ejected into interstellar space from abundant amorphous water-ice-coated cosmic dust produced by supernovae. Possible connections to and hypothetical solutions of several astrophysical and astrochemical problems – dark matter, dark energy, astrobiology, and the RNA world as the origin of life on Earth and habitable exoplanets – were presented. Although the expected density of water nanoclusters expelled to space from cosmic dust should be relatively low – approximately $10^{-6}$ of that of hydrogen – their quantum-entangled, spatially delocalized Rydberg electronic states were shown to make cosmic water nanoclusters a candidate for Rydberg Matter (RM), a transparent – over a broad spectrum – low-density substance occupying space more efficiently than any other form of condensed matter [2]. It was further proposed that the ejected water nanoclusters can vibrationally capture, via the microscopic dynamical Casimir effect [3], the high-frequency virtual photons of zero-point-energy vacuum fluctuations above the nanocluster cutoff vibrational frequency, leaving only vacuum fluctuations below this frequency to be active gravitationally, thereby cancelling the "vacuum catastrophe." Also included were novel speculations about the small cosmological constant, the coincidence of energy and matter densities, the Hubble constant crisis, the role of water as a known coolant for rapid early star formation, and finally how life may have originated from RNA protocells on Earth and exoplanets and moons in the habitable zones of developed solar systems [1]. From the quantum chemistry of water nanoclusters interacting with prebiotic organic molecules, amino acids, and RNA protocells on early Earth and habitable exoplanets, this scenario was argued to be consistent with the anthropic principle that our universe must have properties which allow life, as we know it – based on water, to develop at the present stage of its history. In the present paper, several recent experimental observations that are believed to support this phenomenological model are described. An additional experiment using the James Webb Space Telescope to test the theory further is proposed.

## II. EJECTION OF WATER NANOCLUSTERS FROM AMORPHOUS WATER-ICE

Protonated $(H_2O)_nH^+$ nanoclusters ejected by ion bombardment of amorphous water-ice simulating that on cosmic dust have been observed over a range of n-values [4]. The "magic-number," $n = 21$, $(H_2O)_{21}H^+$ nanocluster (Fig. 1), or equivalent $H_3O^+(H_2O)_{20}$ cluster, is exceptionally stable in a vacuum. It is of possible cosmic importance because it can be viewed as a $H_3O^+$ (hydronium) ion "clathrated" by a pentagonal dodecahedral cage of twenty water molecules [5]. Widespread interstellar $H_3O^+$ has been observed [6]. The infrared signature of $H_3O^+(H_2O)_{20}$ is known [7,8]. For example, cosmic ray ionization of $H_2$ molecules adsorbed on amorphous water-ice-coated cosmic-dust grains can lead to the reaction [9]

$$H_2^+ + nH_2O + \text{grain} \rightarrow H_3O^+(H_2O)_{n-1}\uparrow + \text{grain}. \tag{1}$$

### III. THE ABSENCE OF DARK MATTER IN GALAXY 1052-DF2

The Hubble Space Telescope has revealed a galaxy, NGC 1052-DF2, shown in Fig. 2, approximately seventy-two million light years from Earth, where one can literally see through to other galaxies behind it [10]. It is an ultra-diffuse galaxy, practically as wide as the Milky Way, but contains only roughly 1/200 of the number of stars in the Milky Way. Furthermore, the galaxy contains, at most, only 1/400 of the amount of dark matter that is expected from standard dark matter theory. The ultra-diffuseness of the galaxy also suggests a lack of the cosmic dust characteristic of the Milky Way and most other galaxies. It is proposed here that the absence of <u>both</u> dark matter and cosmic dust in NGC 1052-DF2 is not just coincidence – nor can be justified by some mysterious process removing expected dark matter – but can be explained straightforwardly by cosmic dust as a <u>source</u> of dark matter, as proposed in Ref. [1].

### IV. INTERACTION OF DARK MATTER WITH GALACTIC BARYONIC DARK MATTER

Based on recent extensive studies of distant spiral galaxies, it has recently been discovered there is a direct interaction, in additional to gravitational, between dark matter and the normal baryonic matter in those galaxies [11]. Because the densities of cosmic dust in galaxies track those of the dark matter out to the less-dense regions of the halos, it is suggested here that at least some dark matter is Rydberg matter composed of dust-ejected water nanoclusters that interact both gravitationally and electromagnetically with the normal galactic baryonic matter, in qualitative agreement with the predictions of Ref. [1].

### V. THE BULLET CLUSTER

Gravitational lensing observations of the *Bullet Cluster*, revealing the separation of normal luminous matter and dark matter, have been said to be the best evidence to date for the existence of dark matter [12]. Because much of the Bullet Cluster normal matter is composed of positively charged cosmic dust [13], coulombic repulsion of protonated water nanoclusters expelled from the dust should enhance their ejection by reactions like Eq. (1), explaining the observed separation of normal luminous matter and dark matter.

### VI. CMB BIREFRINGENCE

The recently reported extraction of cosmic birefringence data from the 2018 Planck Cosmic Microwave Background may support "new physics" such as *quintessence* [14]. A more recent paper [15], based on Planck data release 4, agrees with and is believed to be more accurate data-wise than the former publication, providing further evidence for physics beyond the standard model.

Symmetry-breaking terahertz (THz) vibrations of cosmic water nanoclusters, ejected from amorphous water-ice-coated cosmic dust to interstellar space [1], can produce the dipole-moment anisotropy necessary for their birefringence property. Fig. 3a illustrates the 1.8 THz vibrational mode of an electrically neutral water nanocluster, $(H_2O)_{21}H$ after capturing an electron from space into the Rydberg LUMO of $(H_2O)_{21}H^+$ (Fig. 1a). This mode has been shown to be key to the "finely-tuned" catalytic coupling of water nanoclusters with prebiotic molecules necessary for RNA polymerization, and it is practically equal in frequency to the cutoff value, 1.7 THz of vacuum fluctuations proposed to be responsible for dark energy [1]. Symmetry breaking of the water-cluster dodecahedra via the Pseudo Jahn-Teller effect [16] also produces the dipole moments along the nanocluster axes indicated in Fig. 3a, computed by the SCF-Xα cluster density-functional method [17]. Anisotropic dipole moments lead directly to the birefringence of water nanoclusters, as they do for the observed THz-induced birefringence of a water layer [18], which can be viewed as a network of dodecahedral water clusters (Fig. 4) [19]. Thus, the birefringence of cosmic water nanoclusters ejected by cosmic dust can explain, at least qualitatively, the CMB birefringence from a condensed-matter point of view, instead of by nonbaryonic particles such as AXIONS [20,21].

## VII. QUINTESSENCE – DARK MATTER AND DARK ENERGY INEXORABLY CONNECTED

The ancient Greeks had words for it - the *Fifth Element* or *Quintessence*, an invisible material filling all the unoccupied space in the Universe. When the Greek philosophers associated each of the four classical elements - earth, air, fire, and water - with four of the Platonic solids – cube, octahedron, tetrahedron, and icosahedron, they assigned the remaining fifth regular polyhedron, a pentagonal dodecahedron, to the universe in its entirety. Approximately twenty-five hundred years later, in 2003, Jean-Pierre Luminet of the Observatoire de Paris, studying the 2001 WMAP CMB data, proposed that the shape of our universe is indeed a finite dodecahedron [22]. Alas, other astronomers have searched for more evidence to support this aesthetically pleasing scenario but have found none thus far. Nevertheless, the idea that some fifth element or quintessential substance pervades the Universe, whatever its shape, has persisted with a few astrophysicists who believe it is responsible for a time-varying dark energy that propels the measured accelerating expansion of the universe [23]. The standard model of elementary particle physics does not predict the existence of quintessence. The belief in quintessence contrasts with that of most cosmologists, who associate dark energy with a *cosmological constant* - the one originally invented but later rejected by Albert Einstein. However, calculation of the dark-energy density and thus the cosmological constant from quantum electrodynamics yields a value that is $10^{120}$ times the measured value from CMB observations.

In answer to the question, "Is cosmic birefringence due to dark energy or dark matter?" [20], birefringent water nanoclusters ejected from cosmic dust to the vacuum of space as dark RM collectively constitute a dynamical scalar field of quintessence (Fig, 5) – effectively a "dark-matter superfluid" [24]. RM water nanoclusters are quantum-entangled ("bonded") via occupation and "overlap" of their spatially delocalized, high-principal-quantum-number, $n$, Rydberg molecular orbitals. The lowest, $n$ = 1-4, "S,P,D,F" Rydberg orbitals of $(H_2O)_{21}H^+$ are shown in Fig, 1b. The average RM "bond distance" between nanoclusters is approximately1500 nm for principal quantum number $n$ = 100. This is the long-distance analogue of the overlap of water-cluster P$\pi$ molecular orbitals in a liquid-water layer shown in Fig. 4.

It was shown in Ref. [1] that if the energy density

$$\rho_c = \frac{4\pi h}{c^3} \int_{\nu_c}^{\infty} \nu^3 \, d\nu \tag{2}$$

of the high-frequency virtual photons of zero-point-energy vacuum fluctuations is absorbed by the ejected water nanoclusters above their cutoff vibrational frequency $\nu_c$ through the microscopic dynamical Casimir effect [3], as illustrated in Fig. 6, the effectively infinite dark energy density predicted by quantum-field theory is largely cancelled, leaving the finite quantity, Eq. (3) to be identified with the dark energy density

$$\frac{4\pi h}{c^3} \int_0^{\nu_c} \nu^3 \, d\nu = \frac{\pi h \nu_c^4}{c^3} = \rho_{dark}. \tag{3}$$

Ejected water-nanocluster vibrational kinetic energies ½$\dot{Q}^2$ are insignificant compared to their potential energies V(Q), which are raised to the higher-THz-frequency "surface" vibrational modes shown in Fig. 1c by the capture of zero-point-energy vacuum photons illustrated in Fig. 6. Thus, the quintessence scalar field pressure, P = ½$\dot{Q}^2$ – V(Q) becomes more negative with increasing $\nu_c$ and consequently with $\rho_{dark}$. The PLANCK observations have concluded that dark energy presently constitutes 68.3% of the total known energy of the universe, leading to $\rho_{dark}$ = 3.64 GeV/m$^3$ [25]. Therefore, Eq. 3 requires a cutoff frequency of $\nu_c$ = 1.66 THz, which is approximately the average of the cutoff frequencies, 1.5 and 1.8 THz of the positive and neutral water clusters shown in Figs. 1c and 3a. Because $\nu_c$ decreases with increasing water-cluster size (n > 21 in Eq. (1)), according to Eq. (3), a trend toward the ejection of larger water clusters from cosmic dust over time would imply a decrease of dark energy density over time, and therefore a decreasing acceleration of the universe.

## VIII. CONCLUSIONS

Recent empirical evidence that water nanoclusters are ejected from cosmic dust and may constitute Rydberg baryonic dark matter, as originally proposed in Ref. [1], has been presented. This phenomenological theory is further testable. The water clusters discussed have a clear infrared signature [7.8]. The James Webb Space Telescope might be utilized to identify this signature. This theory does not rule out nonbaryonic dark matter, such as WIMPS and AXIONS, although observational evidence for their existence is wanting. Increasing size of prominent ejected water clusters over time implies decreasing $\nu_c$ in Eq (3) and thus decreasing dark energy and acceleration of the universe with time. The ejected cosmic water nanoclusters constitute a dynamical quintessence scalar field and capture the high-frequency virtual photons of zero-point-energy vacuum fluctuations above the water-nanocluster cutoff vibrational frequency. Only vacuum fluctuations below this frequency remain gravitationally active, thereby cancelling the vacuum catastrophe of quantum electrodynamics and inescapably linking dark energy with dark matter. It may be coincidence that the THz cutoff vibrational frequencies of water nanoclusters correspond to the zero-point vacuum energy THz cutoff frequency that produces the dark-energy density of recent cosmological data, but there is no other known Rydberg substance in space for which this is the case. Nonetheless, a common origin of dark matter and dark energy outside the territory of conventional elementary-particle physics, which has thus far not identified conclusively the origins of dark matter and dark energy, is appealing.

**FIGURE CAPTIONS**

**Fig. 1.** Molecular orbital energy levels, wavefunctions, and vibrational modes of the protonated $(H_2O)_{21}H+$ or equivalent $H_3O^+(H_2O)_{20}$ nanocluster. **a.** Cluster molecular-orbital energy levels. **b.** Wavefunctions of the lowest unoccupied Rydberg cluster molecular orbitals. **c.** THz vibrational spectrum. **d.** Lowest-frequency THz vibrational mode and relative vibrational amplitudes. The vectors show the directions and relative amplitudes for the oscillation of the hydronium $(H_3O^+)$ oxygen ion coupled to the O–O–O "bending" motions of the cluster "surface" oxygen ions. The LUMO and higher Rydberg molecular orbitals readily accept an electron, resulting in a neutral cluster (See Fig. 3).

**Fig. 2.** Galaxy NGC 1052-DF2 is missing most, if not all, of its dark matter. Credit: NASA.

**Fig. 3 a.** 1.8 THz vibrational mode of the electrically neutral, symmetry-broken dodecahedral water nanocluster, $(H_2O)_{21}H$ after capturing an electron in the Rydberg LUMO of $(H_2O)_{21}H^+$ (Fig. 1a). The component anisotropic dipole moments (in Debyes) along the nanocluster axes are indicated. **b.** Molecular orbital wavefunction $\Psi$ of the Rydberg LUMO holding an added electron.

**Fig. 4.** The overlap of dodecahedral water-cluster P$\pi$ molecular orbitals in a liquid-water layer. $\Psi+$ and $\Psi-$ represent opposite phases of the wavefunction.

**Fig. 5.** Illustration of the ejection of water nanoclusters from a cosmic dust particle covered by amorphous water-ice. The water clusters can be viewed as constituting a quintessence scalar field. The image can be viewed in 3d stereo with red-cyan glasses. See online animation: https://youtu.be/N3ybQX2i6Mo.

**Fig 6.** Microscopic dynamical Casimir capture of zero-point-energy vacuum fluctuation high-frequency "virtual" photons by THz vibrational states of a cosmic water nanocluster. The water-cluster "mechanical" THz vibrational frequencies are smaller than the optical transition frequencies, leading to the conversion of virtual photons to real ones, Ref [3]. See online animation: https://youtu.be/N3ybQX2i6Mo.

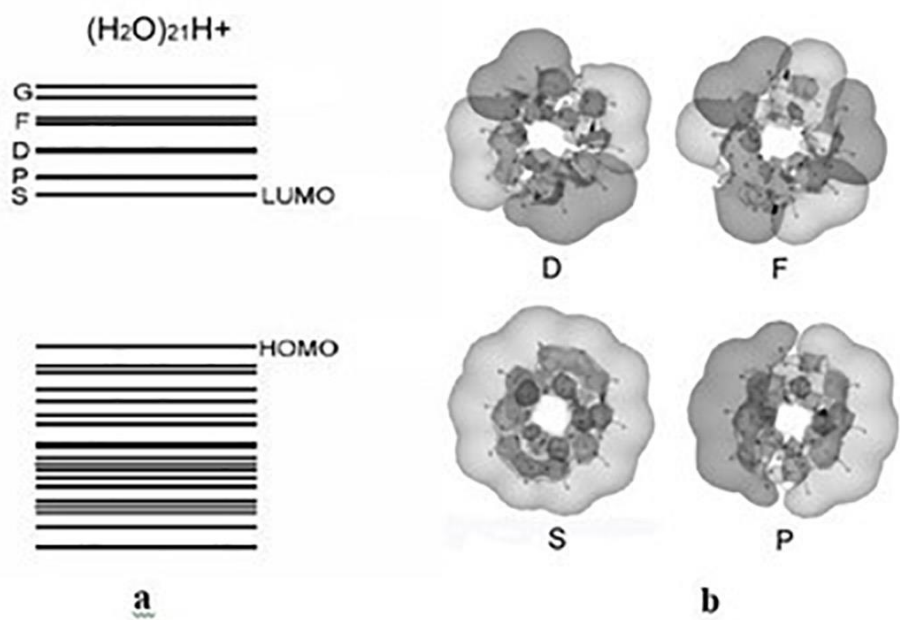
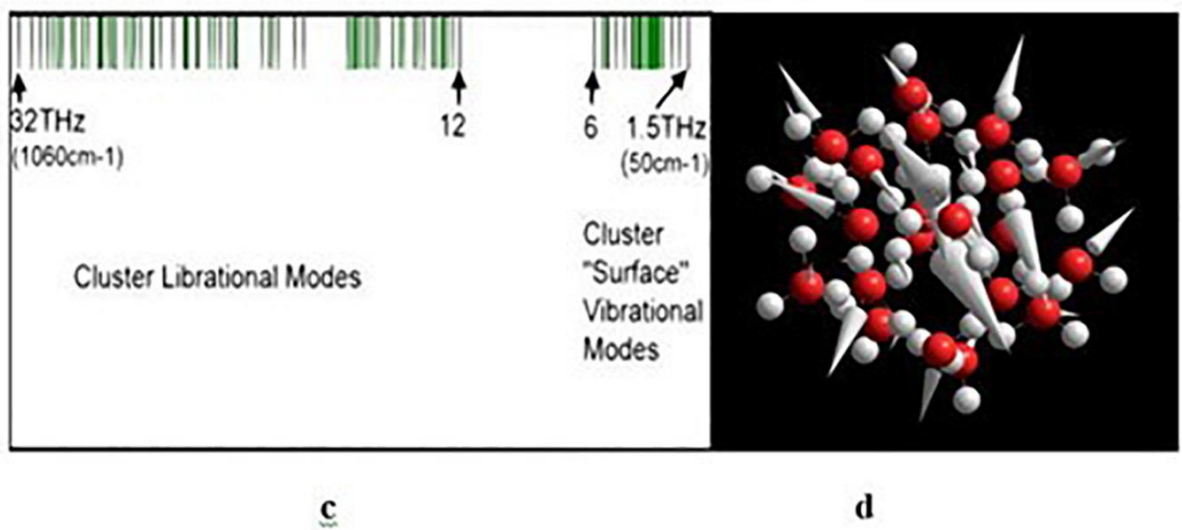

**Fig. 1**

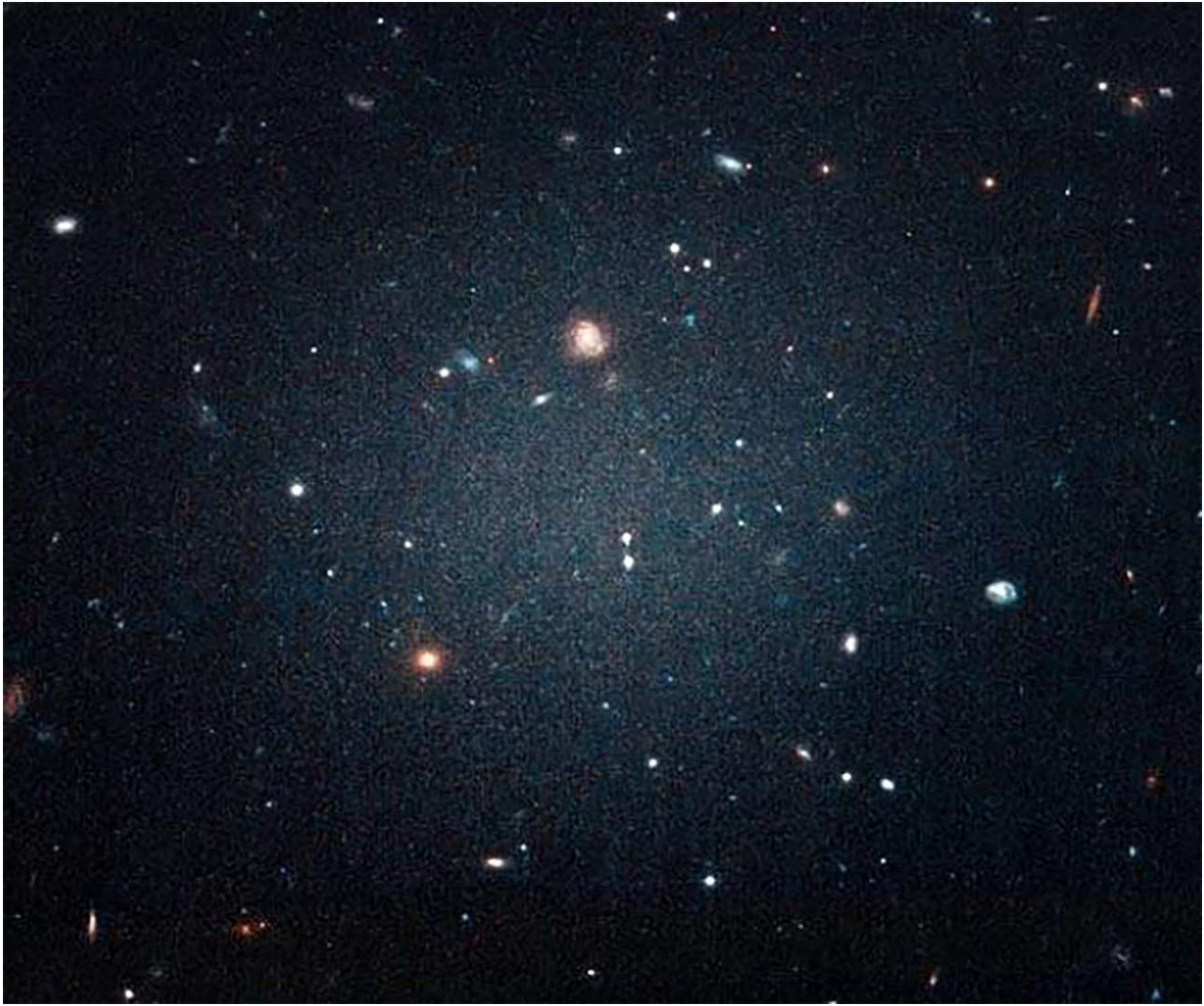

**Fig. 2**

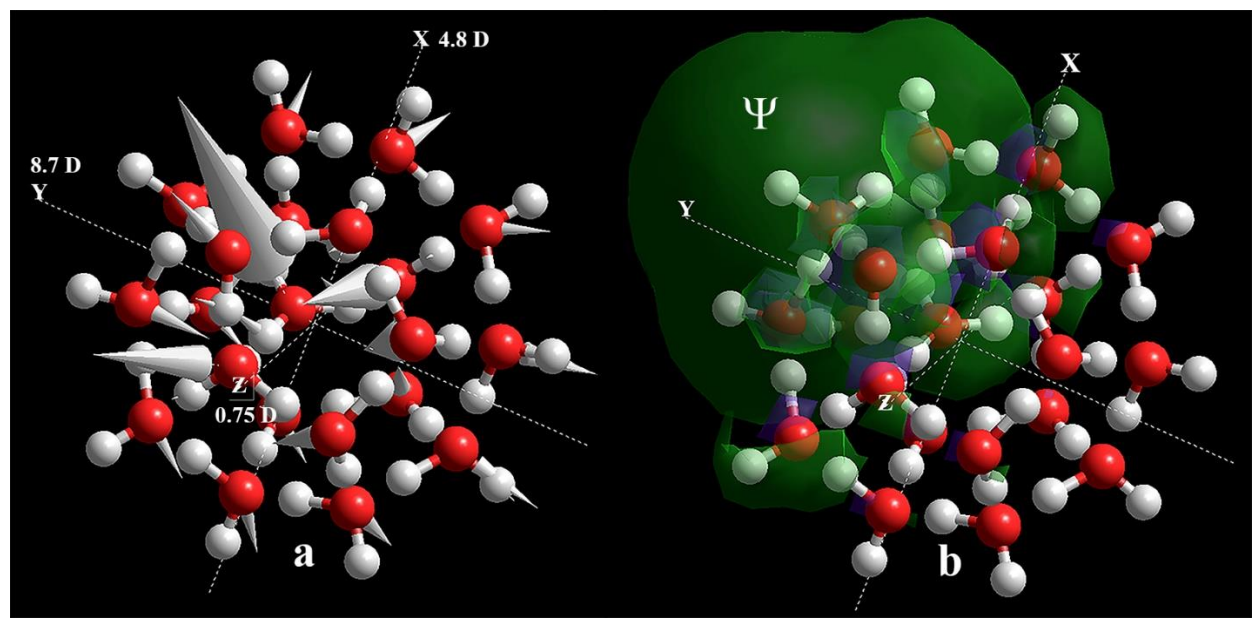

**Fig. 3**

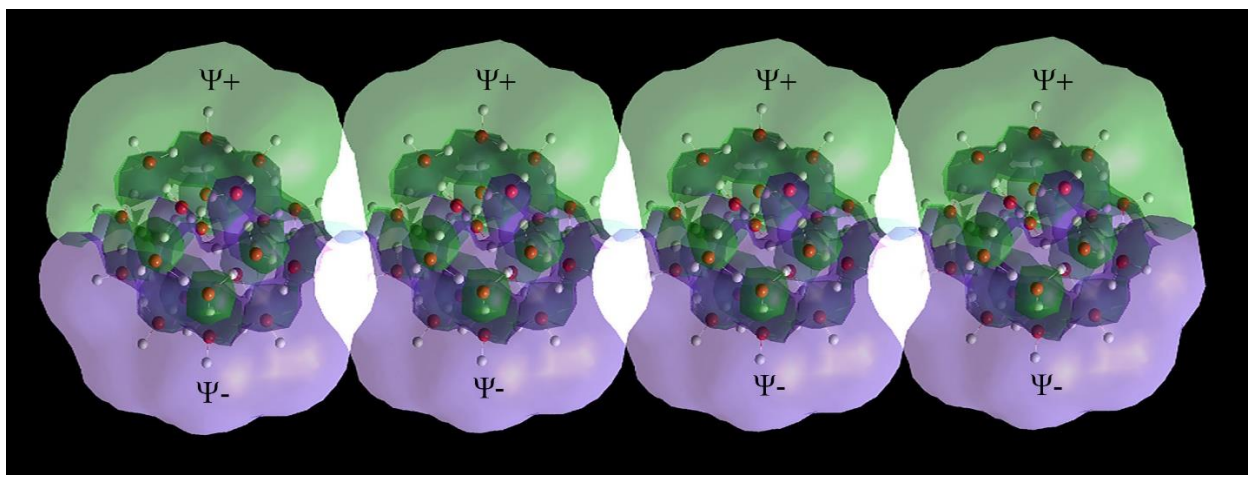

**Fig. 4**

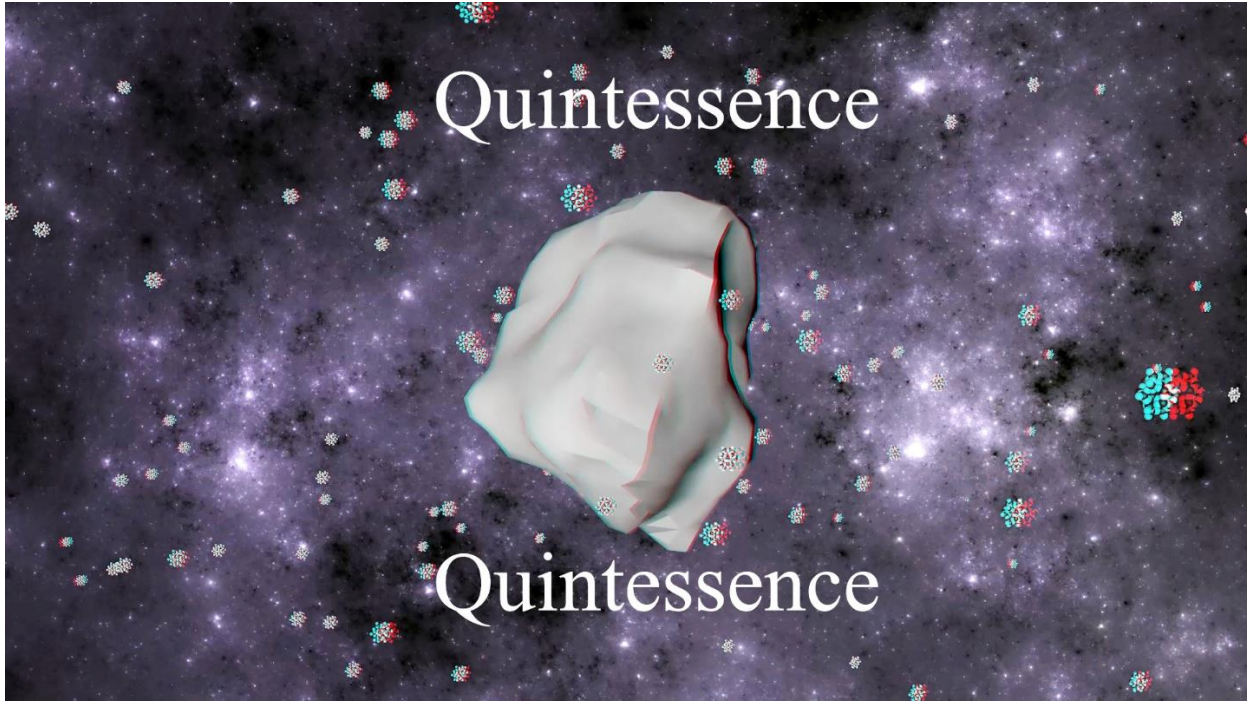

**Fig. 5**

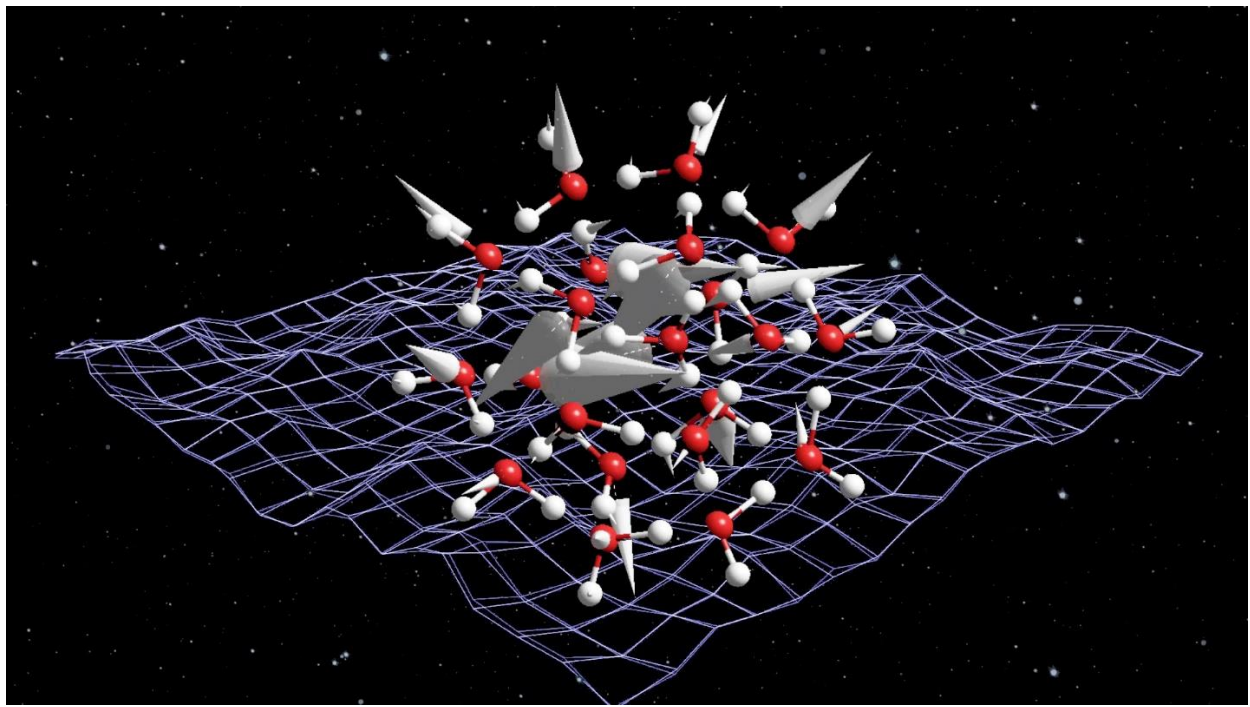

**Fig. 6**